\documentstyle[preprint,aps,eqsecnum]{revtex}
\begin{document}
\draft
\title{Specific plateaus of the quantum Hall effect induced by an 
applied bias.}
\author{A.~Aldea$^*$}
\address{Institut f\"{u}r Theoretische Physik, Universit\"{a}t zu K\"{o}ln,
D-50937,Germany}
\author{P.~Gartner, A.~Manolescu, and M.~Ni\c{t}\~a}
\address{Institute of Physics and Technology of Materials, POBox MG7,
Bucharest-Magurele, Romania}
\date{17.12.1996}
\maketitle
\begin{abstract}
The spectrum and the eigenstates of a finite 2D tight-binding electronic
system, with Dirichlet boundary conditions, in magnetic field and external
linear potential are studied. The eigenstates show an equipotential
character and may cross  the plaquette in the direction perpendicular
to the electric field.  When leads are added to the plaquette,  the channels
carrying the current may be shortcut by equipotentials, resulting in
additional plateaus situated inbetween the usual IQHE plateaus.
This idea is confirmed by a numerical calculation within the four-terminal
Landauer-B\"{u}ttiker approach.
\end{abstract}
\pacs{73.20Dx, 73.23.-b, 73.40.Hm}

The Hofstadter spectrum, characteristic to an ordered plaquette with periodic
boundary conditions along both directions (torus geometry) and a rational
magnetic flux through the unit cell, shows well-defined gaps 
specified by the gap
quantum numbers (s, $\sigma$) satisfying a Diophantine equation \cite{3}.
When  Dirichlet boundary conditions are imposed along one direction
(cylinder geometry), the gaps get filled with edge states. However, as long 
as the density of edge states is small compared to the density of band states, 
the gaps can be clearly observed (except eventually in the center of the 
spectrum) and we shall call them quasi-gaps. In both situations, 
when the Fermi energy lies in a (quasi-)gap, the Hall conductance is 
quantized and can be expressed either as a topological invariant described 
by $\sigma$ or as a winding number of the edge states \cite{1}. Introducing 
now  Dirichlet boundary conditions all around the plaquette the qualitative 
aspect of the spectrum is preserved, although the edge states in quasi-gaps 
are denser. Fig.1 shows  a quarter of the Hofstadter butterfly in this latter 
case.  The usual topological arguments do not hold any more for 
the discussion of 
the Hall effect. We shall make use of a numerical calculation, based on the 
Landau-B\"{u}ttiker formalism with four terminals.  For the ordered plaquette, 
the Hall and the longitudinal resistances still show the typical plateaus of 
the IQHE, their accuracy being determined by the size of the system, 
and by the number of channels supported by the leads (see later in Fig.7).

The energy spectrum may be strongly affected by an applied bias; as expected, 
the tendency is to close the gaps, as it can be seen in Fig.2, where only a 
reminiscent quasi-gap can be observed.  To some extent, this effect is 
similar to that of the disorder, although the mechanisms are different. 
Indeed, with increasing disorder the bands become wider and eventually the 
gaps disappear.  In the disordered case, the presence of the (quasi-)gaps is 
necessary for the existence of the IQHE.  The current is carried by the 
edge states, while the band states become localized, with a fractal 
structure, leading to fluctuations of the Hall resistance \cite{2}.

In this  paper we shall show that, when an external bias is applied, Hall 
plateaus  may exist even in the absence of quasi-gaps.  The eigenstates have 
an equipotential character, i.e. they are localized along the equipotential 
lines of the external potential, and this is sufficient to give rise to   
plateaus and also to new effects at the transition between them.

Let us consider a 2D square lattice with N sites in the $x$ direction
and M sites in the $y$ direction (lattice constant=1).  The one-electron 
problem  defined on this plaquette in the presence of a perpendicular
magnetic field  is described by the following tight-binding
Hamiltonian \cite{3}:
\begin{eqnarray}
&&H^{S}=\sum_{n=1}^{N}\sum_{m=1}^{M} [~\epsilon_{nm} \vert n,m
\rangle\langle n,m\vert 
+ e^{i2\pi\phi m}\vert n,m\rangle\langle n+1,m\vert 
+\vert n,m\rangle\langle n,m+1\vert + h.c.~] \,,
\end{eqnarray}
${\vert n,m \rangle}$ being a set of orthonormal  states localized at the 
sites (n,m), and $\phi$  the magnetic flux through the unit cell measured 
in quantum flux units $\phi_{0}$. The external linear potential is simulated 
by choosing the diagonal matrix elements of the form :
\begin{eqnarray}
\epsilon_{nm}=dV\,(n-N/2)\,,\,\,\,\, n=1,...,N \,.
\nonumber
\end{eqnarray}

Before addressing the problem of transport coefficients, it is useful to 
mention that, according to the invariance properties of the Hamiltonian (1),
the eigenvalues have the obvious symmetries : i) $E_{p}(\phi)= E_{p}(-\phi)$, 
ii) $E_{p}(\phi)=E_{p}(\phi+1)$, and iii) both $E_{p}$ and $-E_{p}$ belong 
to the spectrum.  Therefore the energy spectrum can be reduced to only
a quarter of the Hofstadter butterfly. Of special interest is the spacial 
distribution of the eigenstates on the plaquette, at various energies, for 
$dV\not=0$. An overall view can be obtained looking at the projection of 
all eigenvectors on a line crossing the plaquette along the $x$ direction 
(which is the direction of the applied electric field), i.e.
$\vert\langle n,m~\vert\Psi_{E}\rangle\vert^{2}$ , for $n=1,...,N$ and 
a fixed $m$.  An exemplification is given in Fig.3 for $\phi=0.15$ .

%figure1
%figure2

One has to notice that the localization of the eigenvectors on the plaquette 
depends both on the energy and on the applied potential:

a) At high energies the eigenvectors have a rectangular shape, with three
sides on the borders of the plaquette (at $y=0$, $y=M$, and $x=N$) and the 
fourth on an equipotential line at $x=N-\Delta n$, with 
$\Delta n=\Delta E/dV$, $\Delta E$ being the energy measured downwards,
from the upper edge of the spectrum, at the given flux.
Such edge states are thus shortcut by the equipotentials of the external
field, and may be called 'inner edge states'.   The corresponding
localization probability, $\vert\Psi\vert^{2}$, over the whole plaquette,  
is displayed in Fig.5a.

b) With decreasing energy the fourth side sweeps the plaquette until it 
touches the border at $x=0$, and a true edge state is recovered.

c) At still lower energies a second band of equipotentials appears, which
overlaps with the third and so on, and consequently there are no other
gaps in the spectrum. In the second band, beside eigenvectors of the aspect
shown in  Fig.5a, there is also another type of eigenvector : it follows 
the edges all around the plaquette, but exhibits a ridge that connects the 
opposite sides of the plaquette in the $y$ direction (see Fig.5b). This ridge 
moves with the energy and is also equipotential-type.

d) in the middle of the spectrum the states become blurred and we shall not 
analyze this region.

%figure3

Now, we want to study the influence of the equipotential character of the
eigenvectors on the transport properties of our plaquette. At this point, we 
need to remind of B\"uttiker's approach to the IQHE in multi-lead mesoscopic 
systems. Such a system is characterized by a conductance matrix  
${g_{\alpha\beta}}$, relating the currents  $I_{\alpha}$ to the potentials 
at the leads $V_{\beta}$:
\begin{eqnarray}
&&I_{\alpha}=\sum_{\beta}g_{\alpha\beta}V_{\beta}\,,\nonumber\\
&&\sum_{\alpha}g_{\alpha\beta}=\sum_{\beta}g_{\alpha\beta}=0\,.
\end{eqnarray}
We consider here a 4-lead system ($\alpha,\beta=1,...,4$), of the type
sketched in Fig.4, and we denote by $n=integer$ the number of open channels 
that connect only the nearest leads in the  sense of the current. The
IQHE is generated by the matrix\cite{4}
$$g_{0}=\left(\matrix{
       -n      &0      &0      &n\cr
	n     &-n      &0      &0\cr
	0      &n      &-n     &0\cr
	0      &0      &n      &-n
\cr}\right)\,.$$ 

The longitudinal and Hall resistances are given (in units of $h/e^2$) by
\begin{eqnarray}
&&R_{L}=R_{14,23}=(g_{24}g_{31}-g_{21}g_{34})/D\,,\nonumber\\
&&R_{H}={(R_{13,24}-R_{24,13})/2}\nonumber\\
&&=(g_{23}g_{41}-g_{21}g_{43}-g_{32}g_{14}+g_{12}g_{34})/2D\,,
\end{eqnarray}
where $D$ is a positive 3x3 subdeterminant of the matrix $g_{0}$.
Indeed, for the matrix $g_0$ one gets $R_{L}=0$ and $R_{H}=1/n$.

%figure 4

We shall look now for the effect of the external bias on the conductance
matrix. We anticipate the idea that the ridges of the equipotentials 
(shown in Fig.5a and b) may shortcut at least one of  the channels carrying  
the current around the  plaquette. The result will depend essentially
on the position where the shortcut occurs, namely :

a) If the equipotential sits between the leads no.1 and no.2, then
the lead no.2 becomes connected to the lead no.4 (via the edge opposite
to the leads) and, instead of $g_{24}=0$ as in $g_{0}$, one has $g_{24}=1$. 
The sum-rule (2) will then automatically determine the whole matrix $g_{a}$:
$$g_{a}=\left(\matrix{
	-n     &0      &0      &n\cr
	n      &-n-1   &0      &1\cr
	0      &n+1    &-n-1   &0\cr
	0      &0      &n+1    &-n-1
\cr}\right)\,.$$
b) If the equipotential sits between the leads no.2 and no.3, then
the lead no.3 will be connected to the lead no.4 and, instead of $g_{34}=0$,
one has $g_{34}=1$. The new matrix will be $g_{b}$ :
$$g_{b}=\left(\matrix{
	-n     &0      &0      &n\cr
	n      &-n     &0      &0\cr
	0      &n      &-n-1   &1\cr
	0      &0      &n+1    &-n-1
\cr}\right)\,.$$

Using Eqs.(3) one immediately gets for these two situations:
\begin{equation}
a)~~\,R_{L}=0\,,\hspace{8mm} R_{H}={1\over n+1}\,,
\end{equation}
and
\begin{equation}
b)~~\,R_{L}={1\over n(n+1)}\,,\hspace{3mm} R_{H}={1\over 2}({1\over n}+
{1\over n+1}).
\end{equation}

The result a) suggests that the edge states are not strictly necessary for 
the occurrence of IQHE; the equipotentials which do not touch all edges, but
give the matrix $g_{a}$, are equally good. The result b) shows that, for 
energies corresponding to the equipotential states situated in the middle of 
the plaquette, intermediate plateaus should appear in the Hall resistance 
at 3/4, 5/12, 7/24,..., but also in the longitudinal one at 1/2, 1/6, 
1/12,... .

%figure 5a and 5b

The heuristic results (4) and (5) will be checked numerically in what follows.
In order to calculate the transport coefficients (Hall and longitudinal
resistance) we have to supplement the Hamiltonian (1) with terms describing  
the leads and the contacts. In our tight-binding model, each lead consists
of $n_{c}$ one-dimensional, semi-infinite perfect chains, stuck to the 
plaquette at consecutive sites. The chains may or may not be interconnected 
by a hopping term in the Hamiltonian. In the first situation, the Hamiltonian 
of the lead $\alpha$ reads
\begin{equation}
H_{\alpha}^{L}=t_{L}\sum_{\nu=1}^{n_{c}}\sum_{n\ge 1}[~\vert \alpha,\nu,n
\rangle\langle
\alpha,\nu,n+1\vert+ ~h.c.~] \,.
\end{equation}
Then, the Landauer-B\"{u}ttiker formalism can be used for the calculation
of the conductance matrix $g_{\alpha\beta}$. This may be expressed in terms 
of the scattering amplitude and calculated by the Green function technique:
\begin{equation}
g_{\alpha \beta}~=~{e^2\over h} T_{\alpha\beta}~=~{e^2\over h}\sum_{\nu,\nu^
\prime}~~\vert
t_{\alpha \nu,\beta\nu^\prime}\vert^2,~~\alpha\not=\beta
\end{equation}
where $t_{\alpha \nu,\beta\nu^\prime}$ is the amplitude of the out-wave
through the lead $\alpha$ and the channel $\nu$ when the in-wave
is sent through $\beta,\nu^\prime$. The Green function formalism used for
calculating the transmission coefficients is described in Ref.5.

The numerical results for the Hall and longitudinal resistances, in the 
presence of both magnetic and electric fields, are shown in Fig.6. It is 
known that the regions near the plateau transitions are the most sensitive 
to perturbations and, for instance, in the case of disordered samples, 
mesoscopic fluctuations may appear \cite{6}.  The same regions are influenced
by the external bias, and the new plateaus appear not only in the Hall 
resistance but also in the longitudinal one.  For comparison, the Hall 
resistance in the absence of the bias is shown in Fig.7.  Although the new
plateaus are not very accurate, we want to stress that their values are very 
close to those predicted by Eqs.(4-5). The matrices $g_{a}$ and $g_{b}$ 
correspond to ideal contacts. Although the numerical data show that this is 
not the case, the results prove stable with respect to the deviation from the 
ideal matrices.  The width of the intermediate plateaus increases with the 
applied potential. Indeed, the equipotential line sweeps the space $\delta n$ 
between lead no.3 and lead no.2 in an energy interval $\delta E=dV~\delta n$.
However, when the potential becomes of the order of the bandwidth the 
picture breaks down.

%figure 6

The experimental results of Cobden and Kogan can be  well described 
by the tight-binding model used in this paper \cite{7}, and this is why
we think the effect predicted here, in the presence of an applied bias,  
should also be observable, for samples in which the FQHE is prevented by
disorder or thermal effects.  Since the new plateaus occur at small
magnetic fluxes, we believe they should not be intrinsically related to the 
discrete structure of our plaquette. 

%figure 7

In conclusion, the tight-binding approximation of the 2D lattice gas in
crossed magnetic and electric fields shows specific spectral and transport 
properties in the quantum Hall regime. The electric field proves a useful
tool, suitable to evidence inner edge states.

\vskip 0.2cm
Acknowledgements:
One of the authors (AA) is very grateful to Professor Johannes Zittartz for his
hospitality at Institut f\"{u}r Theoretische Physik der Universit\"{a}t
K\"{o}ln where part of this work was performed under Sonderforschungsbereich
341. The authors thank Professor Martin Zirnbauer and Dr. Martin Janssen
for very useful discussions.

\vskip 0.2cm
\noindent$^*$Permanent address: Institute of Physics and Technology of
Materials, POBox MG7, Bucharest-Magurele,\\
Romania

\begin{figure}
\caption{The spectrum as function of magnetic flux (in units $\phi_{0}$):
Dirichlet boundary conditions along both directions ($N= 64, M=10$)}
\label{fig1}
\end{figure}
\begin{figure}
\caption{The spectrum as function of magnetic flux in the presence
of an applied external potential
($ dV=0.02 ,N=64, M=10 $).}
\label{fig2}
\end{figure}
\begin{figure}
\caption{Eigenvectors projected along a line parallel to the electric field
($dV=0.02 ,\phi=0.15, N=64, M=10$).}
\label{fig3}
\end{figure}
\begin{figure}
\caption{Plaquette with four terminals.}
\label{fig4}
\end{figure}
\begin{figure}
\caption{${\vert\Psi\vert^2}$ on whole plaquette for: a) E=3.1495
and b) E=2.3298 ($dV=0.02, \phi=0.1, N=64, M=10$).}
\label{fig5}
\end{figure}
\begin{figure}
\caption{Hall (full line) and longitudinal (dashed line) resistance as
function of energy in the presence of
external bias : $\phi=0.10, dV=0.02, N=64,M=10$.
The values predicted by Eqs.(4-5) are indicated by horizontal lines.
(resistances are in units $h/{e^2}$)}
\label{fig6}
\end{figure}
\begin{figure}
\caption{Hall resistance (in units $h/{e^2}$) as
function of energy in the absence of
external bias : $\phi=0.10,dV=0.0,N=64,M=10$.}
\label{fig7}
\end{figure}

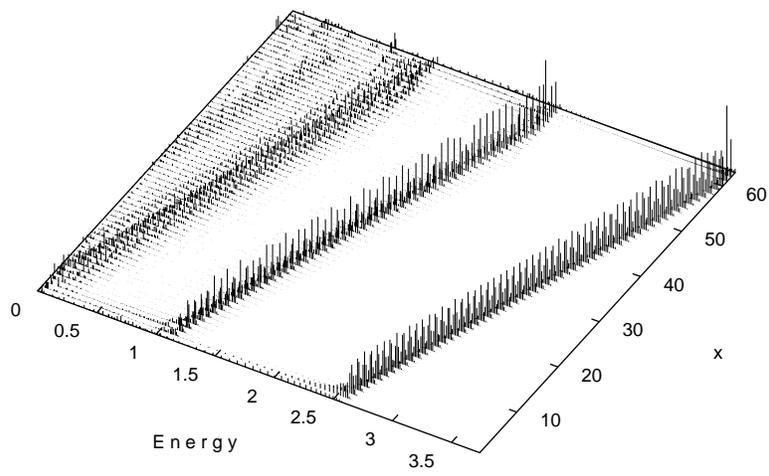
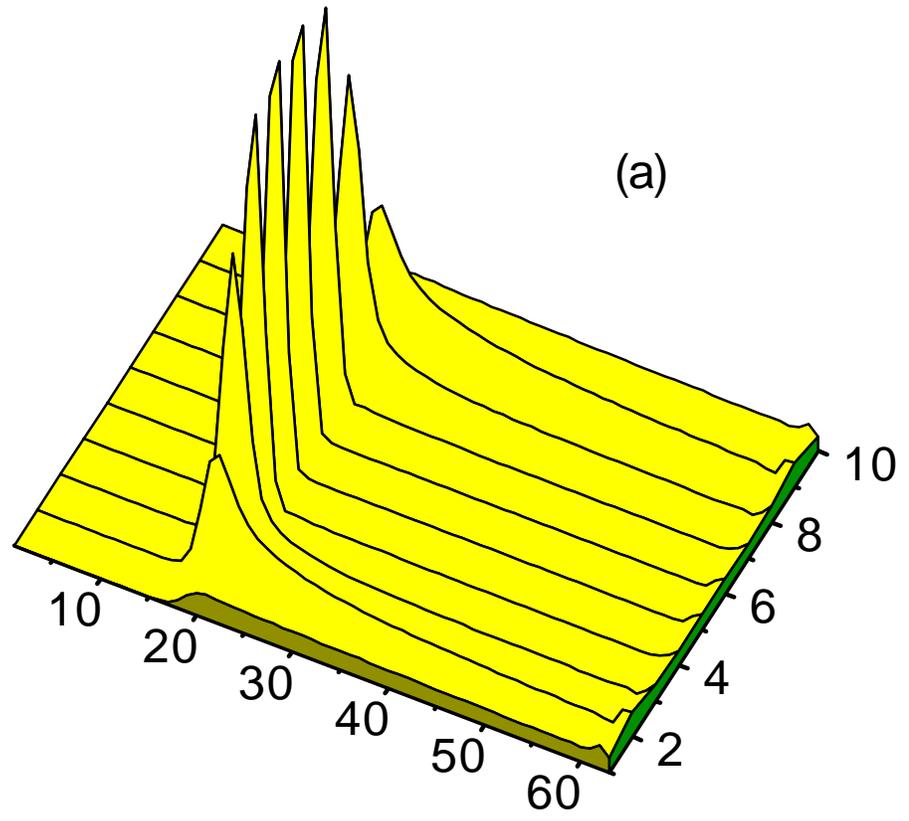
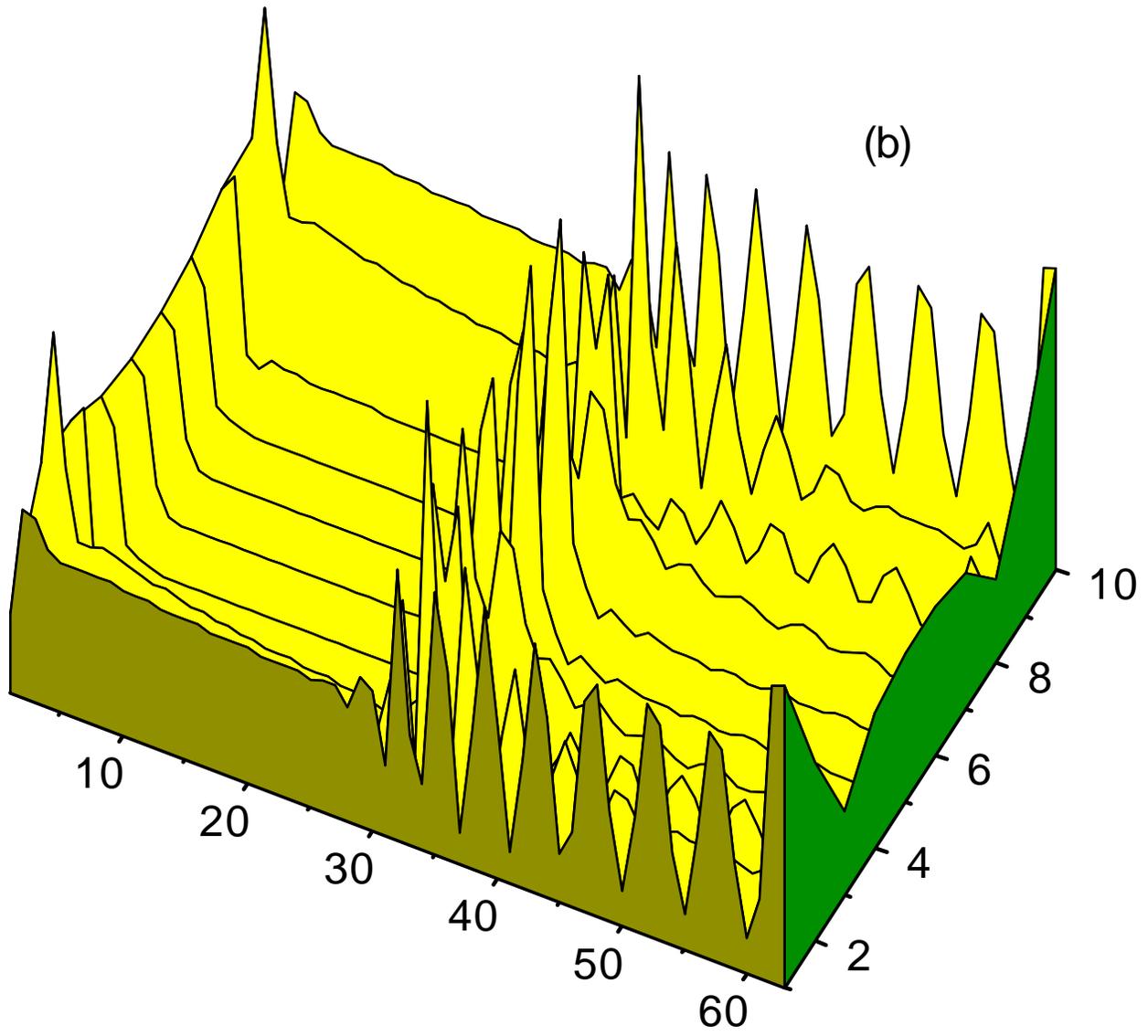
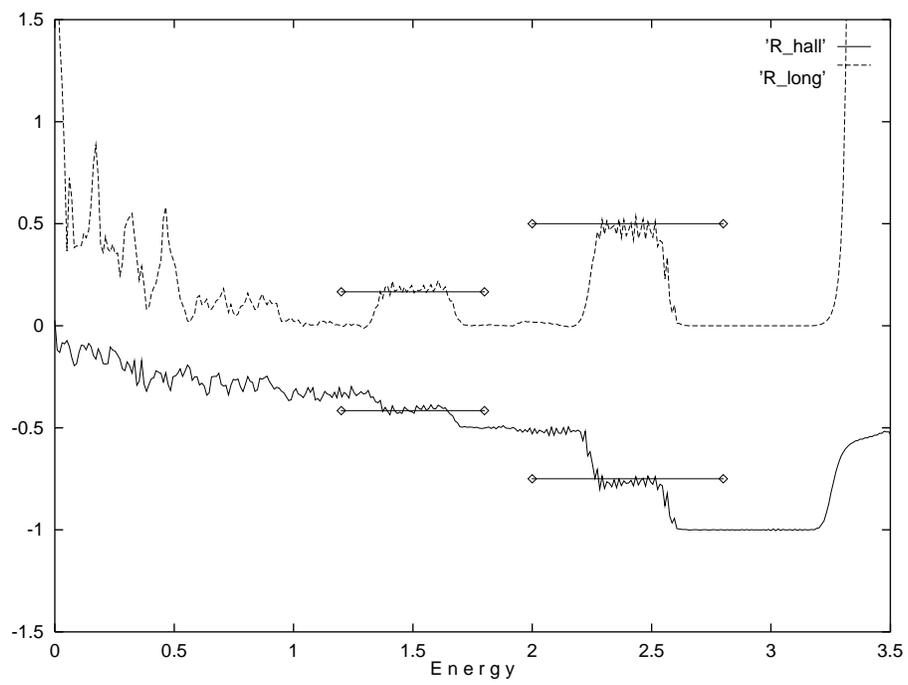
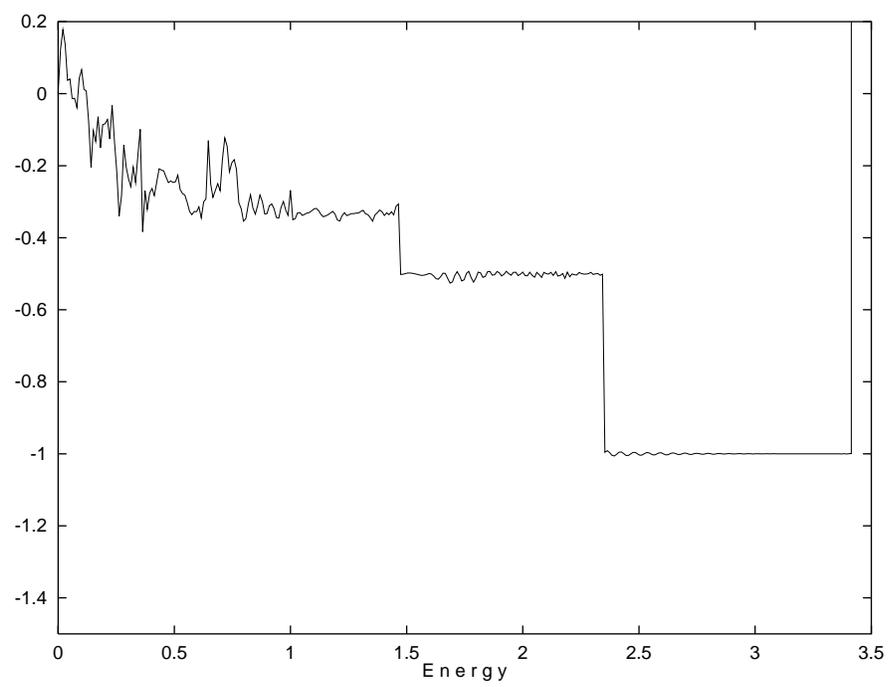
\end{document}